\documentclass[aps,prl,twocolumn]{revtex4}

\usepackage{amsfonts}
\usepackage{amsmath}
\usepackage{amssymb}
\usepackage{bm}
\usepackage{color}
\usepackage{epsfig,graphics,graphicx}

\begin{document}



{\noindent\bf Comment on ``Giant Plasticity of a Quantum Crystal"}

\vspace{6pt}

\noindent
\begin{center}
\small
C. Zhou$^1$, C. Reichhardt$^2$,  M. J. Graf$^2$, J.-J. Su$^3$, A. V. Balatsky$^{2,4}$, and I. J. Beyerlein$^2$
\newline
$^1$Dept.\ of Materials Science \& Engineering, Missouri University of Science and Technology, Rolla, Missouri 65409, USA
\newline
$^2$Theoretical Division, Los Alamos National Laboratory,Los Alamos, New Mexico 87545, USA
\newline
$^3$Dept.\ of Electrophysics, National Chiao Tung University, Hsinchu 300, Taiwan
\newline
$^4$NORDITA, Roslagstullsbacken 23, 106 91 Stockholm, Sweden
\end{center}

In their Letter,  Haziot et al.~\cite{Haziot2013} report a novel phenomenon of giant plasticity for hcp $^4$He quantum crystals. 
They assert that  $^4$He exhibits mechanical properties not found in classical plasticity theory. 
Specifically, they examine high-quality crystals as a function of temperature $T$ and applied strain $\epsilon_{app}$, where the  
shear modulus, $\mu=\tau_{app}/\epsilon_{app}$, reaches a plateau and dissipation $1/Q$ becomes close to zero;
both quantities are reported to be independent of applied stress $\tau_{app}$ and strain $\epsilon_{app}$, 
implying a reversible dissipation process and 
suggesting dislocation motion by quantum tunneling.
At lower $T$,  an increase in $\mu$ and $1/Q$ is found, which is argued to be caused by $^3$He atoms binding to dislocations, thus pinning them and stiffening the solid.

In this Comment, we show that these signatures can be explained with a classical model
of thermally activated dislocation glide without the need to invoke quantum tunneling or dissipationless motion.   
Recently, we proposed a dislocation glide model in solid $^4$He containing the dissipation 
contribution in the presence of other dislocations with qualitatively similar behavior \cite{Zhou2012}.

In Fig.~1(a) we plot our results for both effective shear modulus, $\mu = \tau_{app}/\epsilon_{app}$, 
and the work hardening rate (WHR), $d\tau_{app}/d\epsilon_{app}$, at low ($2.65\times 10^{-8}$) and 
high strain ($1.0\times 10^{-7}$). 
The yield strain at 20 mK is $2.72 \times 10^{-8}$.
In Fig.~1(b) we show the corresponding dissipated plastic energy density 
$dW_p/dt$, which is proportional to $1/Q$ in periodic shear strain measurements.
Notably $\mu$ and WHR exhibit the key features observed in \cite{Haziot2013}, namely they decrease at high $T$.
In addition, our model describes the corresponding dissipation at low strain, see Fig.~1(b), which drops at high $T$, whereas the dissipation remains small and independent of $T$ at high strain.   
This low dissipation arises due to the low Peierls stress that permits dislocations to glide freely. 
However, even in the dislocation freely glide regime
the plastic flow is viscous with finite dissipation.
In Fig.~1(c) we plot $\tau_{app}$ vs.\ $\epsilon_{app}$ at 20 mK to demonstrate that $\tau_{app}/\epsilon_{app}$ is nearly independent of $\epsilon_{app}$ above the yield point; just as is reported in \cite{Haziot2013}, which implies a reversible process.
Clearly the presented results contradict the conclusion of \cite{Haziot2013}, namely that motion in the giant plasticity 
regime is dissipationless. Note that recent torsional oscillator studies no longer show evidence for motion without friction or supersolidity \cite{Kim2012}.

\begin{figure}[th]
\label{fig1}
\includegraphics[width=0.80\columnwidth,angle=0]{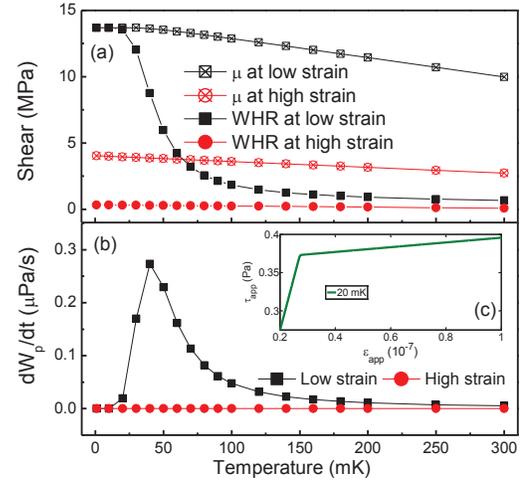}
\caption{(color online) (a) Shear modulus and WHR of solid $^4$He; (b) dissipated plastic energy density;
(c) stress-strain curve at 20 mK.
The model parameters of \cite{Zhou2013} were used.}
\end{figure}

Additional concerns exist with the interpretation of the results of \cite{Haziot2013}:
(a) anisotropy in the dislocation glide does not prove any kind of quantum effect, but can   
arise classically in perfect crystals with specific glide planes and directions of high atomic density;
(b) reversible plastic flow can exist in clean systems with no cross glide \cite{Ziegenhain2009} or  in $ac$ driven systems \cite{Pine2005, Kim2012};  
(c) the low-$T$ dissipation peak is not sufficient evidence for the process of $^3$He atoms binding to dislocations as $T$ is lowered, as our model results show in Fig.~1(b) for a scenario without $^3$He atoms.

In summary, the results in Ref.\ \cite{Haziot2013} can be explained with a classical model of 
dislocation glide and 
without requiring quantum tunneling or dissipationless motion of dislocations or other exotic processes.

\vspace{2pt} 

{\tiny
This work was supported by the U.S.\ DOE at LANL under Contract No.\ DE-AC52-06NA25396 through the LDRD program (C.Z., C.R., I.J.B.)  and the Office of Basic Energy Sciences, Division of Materials Sciences and Engineering (M.J.G, A.V.B).
C.Z. also received partial support from MRC at MS\&T.
Email: zhouc@mst.edu
}

\end{document}